\documentclass[useAMS,usenatbib,fleqn]{mn2e}

\usepackage{times}

\usepackage{amsmath}
\usepackage{amssymb}
\usepackage{upgreek}
\usepackage{graphicx}

\title[Maser Flare Simulations]{Maser Flare Simulations from Oblate and Prolate Clouds}
\author[M. D. Gray, J. Baggott, J. Westlake and S. Etoka]
{M. D. Gray$^{1}$, J. Baggott$^{1}$, J. Westlake$^{1}$ and S.Etoka$^{1}$\\
$^{1}$Jodrell Bank Centre for Astrophysics, School of Physics and Astronomy, University of Manchester,
M13 9PL, UK}
\begin{document}

\date{}

\pagerange{\pageref{firstpage}--\pageref{lastpage}} \pubyear{2018}

\maketitle

\label{firstpage}

\begin{abstract}
We investigated, through numerical models, the flaring variability that may arise from the rotation of maser
clouds of approximately spheroidal geometry, ranging from strongly oblate to strongly
prolate examples. Inversion solutions were obtained for each of these examples over
a range of saturation levels from unsaturated to highly saturated. Formal solutions
were computed for rotating clouds with many randomly chosen rotation axes, and corresponding
averaged maser light curves plotted with statistical information. The dependence of
results on the level of saturation and on the degree of deformation from the spherical
case were investigated in terms of a variability index and duty cycle. It may be
possible to distinguish observationally between flares from oblate and prolate
objects. Maser flares 
from rotation are limited to long timescales (at least a few years) and modest values
of the variability index ($\lesssim 100$), and can be aperiodic or quasi-periodic.
Rotation is therefore not a good model for H$_2$O variability on timescales of weeks to
months, or of truly periodic flares. 
\end{abstract}


\section{Introduction}
\label{intro}

This work is one of a series of publications that describe results of computational
models of flares and variability in astrophysical masers arising from various
possible mechanisms. The first paper in the series, \citet{2018MNRAS.477.2628G}, hereafter Paper~1,
considered variability due to rotation of a model cloud that had only small departures
from spherical symmetry. After a brief general introduction, the present work will
concentrate on the characteristics of flaring caused by the rotation of approximately
spheroidal clouds of varying eccentricity.

Much recent observational interest has centered on correlations between infra-red
variability in the continuum of young stellar objects (YSOs) and maser flares. Correlations
of this type are particularly strong for flares in the 6.7-GHz maser transition of
A-type CH$_3$OH. For example, in the YSO S255~NIRS~3, a 6.7-GHz flare is convincingly
ascribed by \citet{2017A&A...600L...8M} to an infra-red burst due to an accretion event 
associated with the host YSO. A reported 5-month delay between the infra-red and maser
events, coupled with interferometric measurements of the maser and continuum positions
requires information transfer at $>$6000\,km\,s$^{-1}$: too fast for any
mechanical transfer by shock waves that would not also destroy the maser molecules.

In the source NGC 6334I, continuum infra-red variability has been linked to maser
flares in the three species OH (6 transitions), H$_2$O (22\,GHz) and CH$_3$OH (3 transitions)
\citep{2018MNRAS.478.1077M}.
In this source there are various delay times associated with the different distances
to regions of a VLBI map occupied by masers of each species. The maser flares were
linked in this case to infra-red radiation produced by FU~Ori-type accretion events in
a massive protostar, or YSO,  underlying the continuum source MMB1. It is interesting that some
methanol 6.7-GHz spectral features were stable throughout the flare, whilst the flaring
features occupied a new, redder, velocity range, offset by several km\,s$^{-1}$ from the
main pre-flare spectrum.

While there are many instances of CH$_3$OH maser flares that are convingingly linked
to infra-red variability in nearby continuum sources, the evidence for H$_2$O maser
flares is more mixed. In the NGC~6334I source \citep{2018MNRAS.478.1077M}, water masers
appear to share the radiatively driven variability of the other species. In some other
sources, this may also prove to be the case: examples include IRAS18316-0602, where
the H$_2$O flare appears to follow a brightening of a 2MASS source \citep{2017ATel11042....1A,2017ATel10788....1S},
although a shock-driven variation in the pump rate has also been suggested
\citep{2018ARep...62..213L}. In the IR source IRAS~16293-2422, H$_2$O maser flares have been
more decisively linked to motions of shocked gas \citep{2016ARep...60..730C} on the basis of
the spatial distribution of flaring objects along AU-scale chains that contain
velocity gradients. Spectral
shifts in velocity were observed as each successive object was shocked.

Shocks heat and compress gas, but do not guarantee the generation of rotational
motion. However, given the very high Reynolds numbers in most interstellar flows,
it is highly likely that post-shock gas will become turbulent in most shock-driven
maser sources. Turbulent motion in turn drives the development of rotating structures
and eddies that may have very small high-density cores (see for example H$_2$O masers
in Cep~A \citep{2018ApJ...856...60S}. Most
22-GHz H$_2$O masers are assumed to be pumped by a mainly collisional process
following shock heating of the H$_2$O-bearing gas \citep{1989ApJ...346..983E,1996ApJ...456..250K}, so
the rotation model of flares is probably more applicable to water masers than, for
example, Class~II methanol masers that are pumped mostly by infra-red radiation, with
a consequent correlation of maser variability with that of the pumping radiation. 
Water maser flares are also more extreme in variability index so, if a rotation model
is to be considered, it is very important to study significantly non-spherical
clouds. Amplification by a rotating cloud in the line-of-sight has been specifically
considered for a 22-GHz H$_2$O maser flare in W49N \citep{1998ApJ...509..256B}, with a
surprisingly good fit to flux density, line width and centroid velocity via a simple
model. That model, however, considered only an unsaturated rotating cloud between
the observer and an already saturated maser source.

In Paper~1, we described a new 3-D code, specifically written to model maser sources, 
based on clouds of irregular shape. Saturation
effects were treated in a self-consistent manner within the approximations
of the model. The main approximations in Paper~1 were the use of a 2-level model with
phenomenological pumping, an assumption of complete velocity redistribution (CVR) within
the molecular response, an internally uniform cloud (or computational domain), and an
approximately spherical distribution
of triangulation nodes. In the current work, we lift the last of these restrictions in order
to study the behaviour of strongly non-spherical clouds, noting that the finite-element
discretisation of the domain will always lead to small asphericities, even if the node
positions are selected from a spherical distribution.

The scientific driver for the  current works is 
the need to model significantly larger flare amplitudes,
observed from some masers in massive star-forming regions, 
than could be typically produced by the approximately spherical model in Paper~1 that 
yields ratios of maximum to minimum flux density at line centre of $\sim$3. If
flare amplitude is the only consideration, then, for example, the monitoring data in
\citet{2004MNRAS.355..553G} shows that 68 of 372 spectral components, drawn from 54 
6.7-GHz methanol maser sources, have a variability index of $<$3. We note that the variability
index in \citet{2004MNRAS.355..553G} is not simply the maximum to minimum flux density
ratio in a chosen channel, but the point is that a substantial fraction of
the observations cannot be explained in terms of a rotating pseudo-spherical cloud.

When considering 22-GHz H$_2$O masers, their variability behaviour is apparently more
extreme: 43 sources were studied over 20\,yr, with a cadence of 4-5 observations per
year \citep{2007A&A...476..373F,2007IAUS..242..223B}. A variability index used in
\citet{2007IAUS..242..223B} is again defined differently from anything used so far but,
judging from their Fig.~1b, almost all of the 43 sources are
more strongly variable than the model from Paper~1. The issue of multiple definitions
of the variability index is perhaps problematic: we note
that three different variability indices have
so far been introduced, which is potentially confusing. We discuss these definitions
further in Section~\ref{ss:formal}.

\section{Modifications to the Model}
\label{s:model}

The model, including triangulation, discretization, solution for the nodal populations, formal
solution for the specific intensities seen by a distant observer, and the effects
of rotation, remains largely the same as in Paper~1. The obvious difference in the
current work is an additional shaping of the cloud, so that the point distribution
that forms the domain becomes approximately either a prolate or an oblate spheroid.
The shaping operation was inserted between the steps of point generation, into a
volume-weighted spherical distribution as in Paper~1, and triangulation into finite
elements. The shaping algorithm used in the current work was,
\begin{equation}
(x^2 + y^2) e^\Gamma + z^2 e^{-2\Gamma} = 1,
\label{eq:shaping}
\end{equation}
where $\Gamma$ is the deformation factor, a new input parameter of the code. If
$\Gamma < 0$, coordinates in the $xy$-plane are stretched, whilst the $z$-coordinate
is compressed, leading to an oblate spheroidal cloud. Positive values of $\Gamma$
generate prolate clouds; if $\Gamma = 0$, the original approximately spherical
point distribution is preserved. The deformation in eq.(\ref{eq:shaping}) also
conserves the volume of the domain \citep{suganobook}.

The code in Paper~1 ran on a single processor, but the version used in the present
work was adapted to take advantage of multiple-processor architecture by using
the OpenMP application program interface\footnote{https://www.openmp.org}. Use
of OpenMP allowed the parallel operation of 2-12 processors on desktop machines,
and up to 16 processors on the {\sc coma} parallel machine at the Jodrell Bank Centre for
astrophysics.

In the course of this work, the Levenberg-Marquardt algorithm used in Paper~1 was replaced
with a non-linear Orthomin(K) algorithm for the most time-intensive 
task of computing population solutions.
The implementation of Orthomin(K) was based on \citet{2001ApMCo.124..351C} with an
initialization method based on the two-point step size gradient method
\citep{2002ComOA..22..103D,1988IMAJN...8..141B}. The Orthomin method was found to be
less susceptible to convergence problems with increasing optical thickness 
(convergence to an accuracy of 10$^{-8}$ at depth
multipliers of 20 or more is easily achievable with Orthomin, as opposed to 
significant problems at 13.0-13.5 with the Levenberg-Marquardt scheme). Perhaps a
more significant advantage for larger domains is that the Orthomin method does not need
to compute a Jacobian matrix, resulting in many fewer function calls.

\section{Domains}
\label{s:domains}

The prolate and oblate domains were generated by applying distortions based on
eq.(\ref{eq:shaping}) to the same original spherical distribution of points. The
distortions were applied before DeLaunay triangulation to form finite elements. All
domains were generated from the same sequence of 300 points, of which 234-249, dependent on $\Gamma$, 
survived the triangulation process to be incorporated into 
the final domain. A variety of distortion factors were
applied, from mild ($|\Gamma | = 0.1$) to strong ($|\Gamma | = 0.6$). Control results
were also obtained for a pseudo-spherical cloud with the same number
of nodes and zero distortion.

\begin{figure}
  \includegraphics[bb=60 115 595 800, scale=0.65,angle=0]{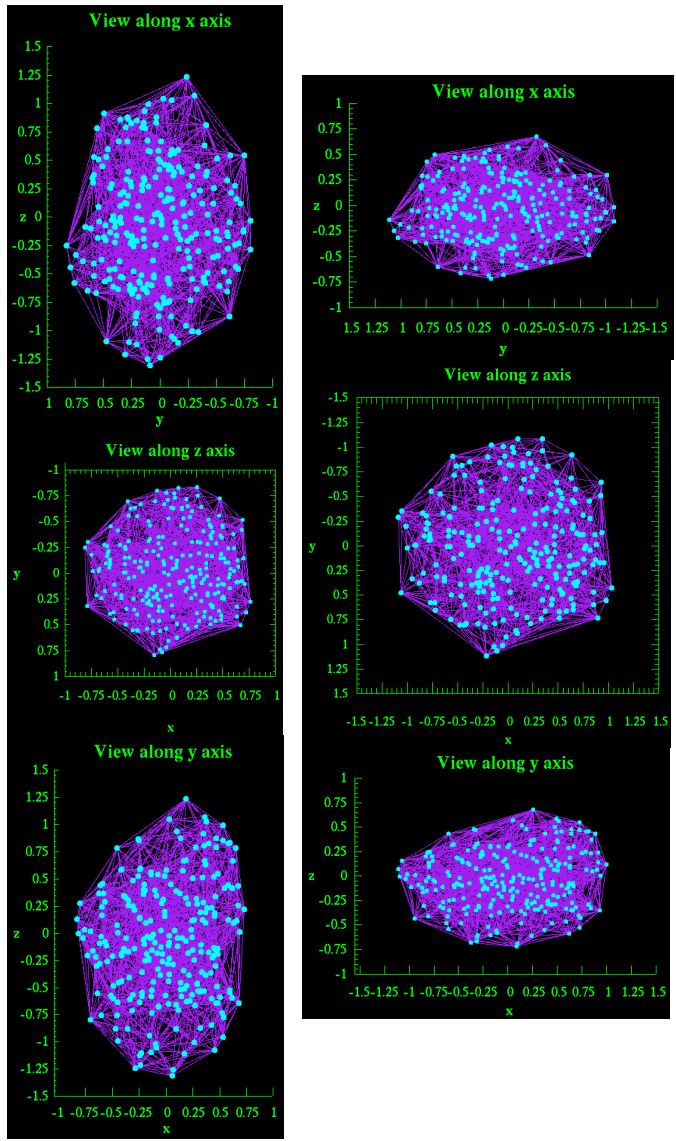}
  \caption{The left-hand panel shows three views of a prolate domain with
$\Gamma = 0.3$; the right hand panel shows the same views of an oblate
domain with distortion $\Gamma = -0.3$. Cyan symbols mark nodal positions, and
purple lines, the element edges.}
\label{f:domains}
\end{figure}

In Figure~\ref{f:domains} we show three views of a prolate domain with $\Gamma = +0.3$ (left-hand
panel) with the observer's viewing direction as marked. In all cases, the view along
a specified cartesian axis is in the positive direction. The right-hand panel shows three
views of an oblate domain with $\Gamma = -0.3$, also with observer's viewing directions as marked.

Oblate clouds are likely to result from the shock compression of an initially spherical
cloud, with the cloud flattened parallel to the shock front. An alternative generation
mechanism is that of an initially spherical cloud with angular momentum, expanding into its surroudings and
becoming rotationally flattened.

It is arguably more difficult to generate a prolate cloud, but one possibility is
the expansion of an initially spherical cloud in a medium threaded by a magnetic field,
a mechanism suggested for the shaping of the envelope of the OH/IR star
OH26.5+0.6 \citep{2010MNRAS.406.2218E}.
If the cloud medium is frozen to field lines that are dynamically dominant, then the
cloud will be able to expand easily parallel to the field lines, but only with
great difficulty perpendicular to the field. Certain  hydrodynamic and magnetohydrodynamic
instabilities can also lead to elongated cloud structures, for example the Rayleigh-Taylor
instability and the sausage instability.

The ray-tracing algorithm was as used in Paper~1, so that, in the population (or inversion) solution,
1442 rays were traced from points on a celestial sphere-style source to every target node
of the domain.
The increased number of nodes compared to Paper~1, meant that somewhat larger total numbers of rays and 
saturation coefficients were generated. The celestial sphere source had the same specific
intensity as in Paper~1: $i_{BG}=1.0\times 10^{-5}$, where $i_{BG}=I_{BG}/I_{sat}$, and $I_{sat}$
is the saturation intensity of the maser.

\section{Results}
\label{s:results}

First, we comment on the population solutions calculated
for selected prolate and oblate clouds, with $|\Gamma | =0.3$, with particular emphasis on the differences 
from the pseudo-spherical cloud
solution in Paper~1. Secondly, we discuss in detail the formal solutions, with simulations of
rotation, that were used to generate light-curves for the clouds, as seen by a remote observer.
Unlike the work in Paper~1, the present work allows the rotation axis of the domain to be
selected independently of the long axis. In all cases, the observer's plane is defined to be
perpendicular to the rotation axis.

\subsection{Nodal Solutions}
\label{ss:nodalsol}

Nodal solutions were generated in the same manner as in Paper~1: an initially very optically
thin solution was obtained, and this solution (and at higher depths, several previous solutions)
were then used as, or to predict, a first guess at an optically deeper solution. This process was 
originally continued
until progress became very slow, at depth multipliers in the range 12.0-13.5. However,
the adoption of the Orthomin(K) algorithm (see Section~\ref{s:model}) has enabled 
us to consider results up to depth multipliers of at least 20.
It should be noted from Fig.~\ref{f:domains} that
the actual maser depth implied by a given value of the depth multiplier is not the same as
in the pseudo-spherical model in Paper~1. Whilst the shorter axis or axes still have a maximum
chord length of approximately 1.8, the stretched axis, or axes, now have larger maximum chords
of approximately 2.7 in the prolate case and 2.5 in the oblate case. A fixed value of the
depth multiplier therefore corresponds to a generally higher maser depth than in the spherical
case, with consequently greater optical thickness along some ray paths.

\begin{figure}
  \includegraphics[bb=0 170 595 790, scale=0.80,angle=0]{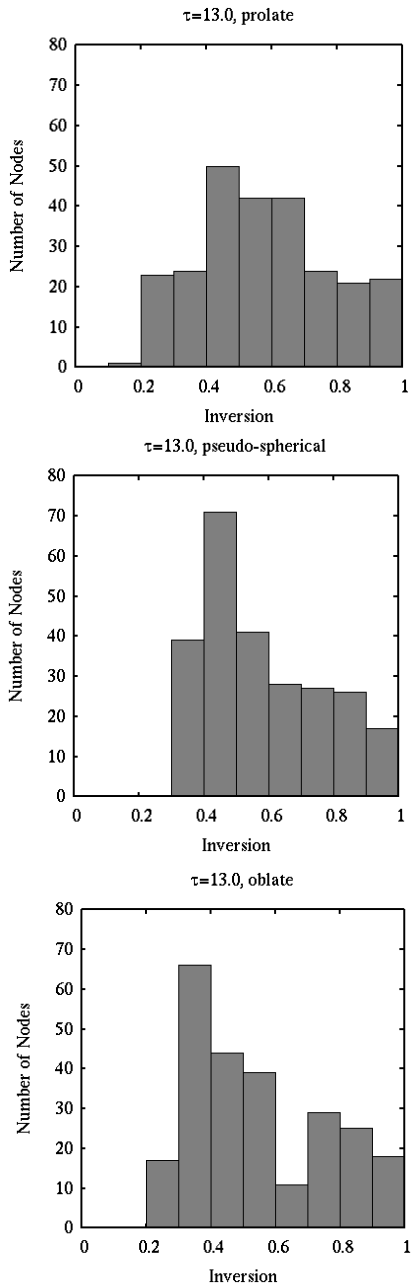}
  \caption{Histograms of the distribution of nodes between bins of fractional inversion
with width 0.1. All three plots are at a depth multiplier of 13.0. The prolate model is
at the top, the middle panel shows the pseudo-spherical model, and the oblate model is
at the bottom.}
\label{f:histos}
\end{figure}

Population solutions are not the main focus of this work, but for completeness we show
in Fig.~\ref{f:histos} the distribution of the nodal inversions amongst 10 
fractional bins for the prolate and oblate domains from Fig.~\ref{f:domains} with
the depth multiplier of 13.0. For comparison, we also show the histogram for the
undistorted control case. In all cases,
the background radiation was uniform at a level of 10$^{-5}$ with respect to the saturation
intensity.

In the prolate case, the most saturated node is at position $(0.2059,-0.2410,1.2362)$, placed
well up the $z$-axis from the origin that is at the centre of the domain. This node is at a distance
of 1.2762 from the origin, and is fairly easy to identify in both the top and bottom 
left-hand panels of Fig.~\ref{f:domains}. This node has only 0.196519 of its original inversion
remaining. By contrast, the most saturated node in the pseudo-spherical model is only
at a distance of 0.9938 from the origin, is significantly less saturated (0.31821 of the
inversion remains) and its position, $(-0.4872,0.6433,0.5801)$, favours no particular axis.
In the oblate case, the most saturated node is at $(-1.0880,-0.2870,0.0736)$, lying close to
the $xy$-plane, and at a large distance from the origin (1.1276). As with the other models, the
least saturated node is again close to the origin of the domain. All these results are 
consistent with expectations, and support the premise that the model is working correctly.

In Fig.~\ref{f:histos}, all the domains show significant evidence of saturation:
the bin with the greatest number of nodes corresponds to a surviving
inversion of 0.35 or 0.45. However, in both the prolate and
oblate cases, the are more examples of highly saturated nodes, with the bin centered on
an inversion of 0.25 well populated, and one node in the 0.15 bin in the prolate case.
It can be argued that  these results are as expected, since both the prolate and oblate models
contain longer chords than are possible in
the pseudo-spherical model, and a chord of length $\sim$2.5 is possible in the prolate
model (see Fig.~\ref{f:domains}). All models continue to have a largely unsaturated
core of $\sim$20 nodes (most right-hand bin in Fig.~\ref{f:histos}). It is also perhaps
worth mentioning the more bi-modal nature of the oblate distribution, with a rather low
number of nodes in the bin centered on an inversion of 0.65.

\subsection{Formal Solutions and Rotation}
\label{ss:formal}

In the work discussed here, formal solutions were obtained by solving the maser radiative
transfer equation, as in Paper~1, with known nodal inversions. In each formal solution, the
rays pass through the domain towards a specified observer's position. We note that, in the
case of the prolate cloud, an observer's view along the long axis of the domain is strongly
privileged, and expected to generate significantly stronger emission than any other viewpoint.
Although a long axis can be determined in the other models, it will have many close rivals
close to the $xy$-plane in the oblate model, and in arbitrary directions in the pseudo-spherical
model.

As the current work deals with temporal variation resulting from cloud rotation, all the
formal solutions were performed in the rotating frame of the cloud, which had no internal
motion. Doppler corrections were then applied to the frequencies of the emitted rays to
render these frequencies correct for a fixed observer viewing a rotating cloud. As in Paper~1,
the dimensionless rotation rate corresponded to 1.162 Doppler widths unless otherwise stated.

\begin{figure*}
  \includegraphics[bb=0 155 595 785, scale=0.85,angle=0]{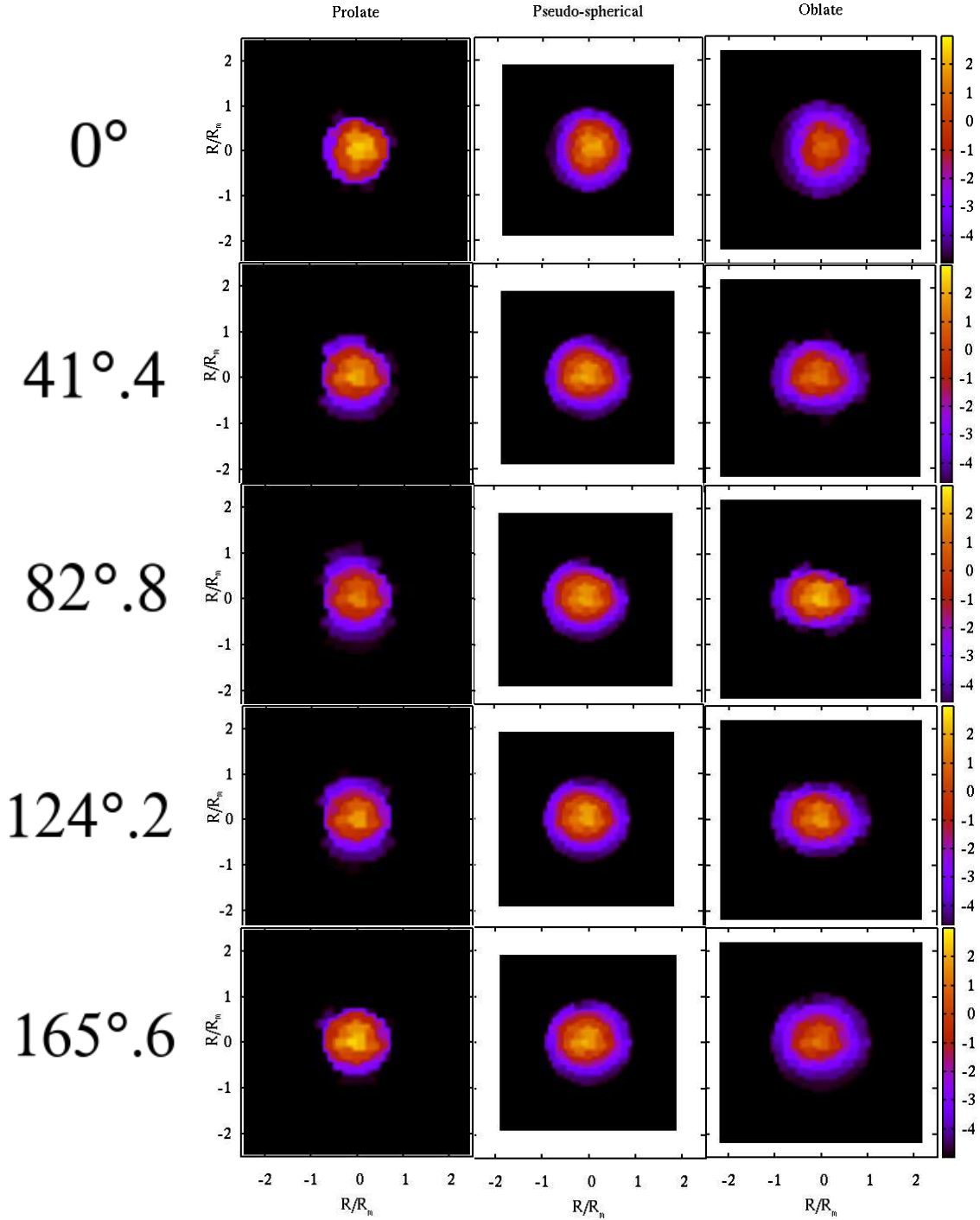}
  \caption{Images of the prolate domain (left-hand column), the pseudo-spherical
domain (centre column) and the oblate domain (right-hand column). The five rows correspond
to views with the observer's position in the $yz$ plane at the angles marked, as measured
from the North pole. The colour scale is the base-10 logarithm of the specific intensity in
units of the saturation intensity.}
\label{f:maps}
\end{figure*}

The initial observer's plane was chosen to be the $yz$-plane, and in Fig.~\ref{f:maps} we show
a specific intensity (maser brightness) map at five positions in this plane for the
same three domains (prolate, pseudo-spherical and oblate) as used for Fig.~\ref{f:histos}.
The colour palettes are the same in all panels of Fig.~\ref{f:maps} to make the brightness variations
between different viewpoints, and between the different domains, as clear as possible.

All domains are shown to the same spatial scale, which leads to a slightly different background
field in each case, since this has a size proportional to the long-axis of the domain in the sky-plane.
In the case of the prolate cloud (left-hand
column of Fig.~\ref{f:maps}, the top-most
image has the observer viewing the North pole of the domain, almost along its long axis, and
there is a small, but substantial,
area that saturates the upper end of the colour table (bright yellow) representing specific
intensities $>$1000\,$I_{sat}$. In fact, the maximum intensity in this map is 1209$I_{sat}$.
The image at 82\degr.8 in the prolate column is the closest to its thinnest
possible presentation to the observer; here the peak brightnesses are only of order $100I_{sat}$. The
brightest ray in fact has a specific intensity of 102$I_{sat}$.
Therefore, rotation of the prolate domain, with
this level of saturation, in a viewing plane that contains the long-axis is
expected to show a maximum brightness variation of order of 10-20.

The pseudo-spherical cloud (middle column in Fig.~\ref{f:maps}) is expected to exhibit a
variability of similar amplitude to the cloud in Paper~1, that is a ratio of maximum to
minimum of $\sim$3. This is borne out by the sequence of maximum intensities, working
down the middle column of 285, 411, 307, 393 and 511 times the saturation intensity. Clearly 
the expected 
amplitude of variation from the prolate
cloud is considerably larger.

In the case of the oblate cloud (right-hand column of Fig.~\ref{f:maps}), the top-most
image is the thinnest possible view of this cloud, viewed towards the North pole,
or anti-parallel to the $z$-axis. As expected,
the maximum brightness of 24.5$I_{sat}$
is much lower than for the same view of the prolate cloud (column
1, angle 0\degr) at 1209$I_{sat}$. The
middle figure of the right-hand column (at 82\degr.8) shows the oblate cloud closest
to its major plane, the $xy$-plane. The long
axis of the cloud is expected to be in, or close to, this plane, so
high brightness is expected. The largest specific intensity in this map was found to be
771 times the saturation value, so large amplitude variability is expected.
In fact, this example is even more extreme, based on
the ratio of maximum to minimum specific intensities found,
than for the prolate case of the same $|\Gamma|$, axis of rotation and
level of saturation.

In this work, we use from now on a definition of the variability index based on the ratio of the
brightest specific intensity found at maximum light to the brightest found at
minimum light. This definition, and two others mentioned in the Introduction, are
set out mathematically in Appendix~\ref{a:indices}. Our variability index specifically corresponds
to eq.(\ref{eq:appA1}). The \citet{2007IAUS..242..223B} index is the 
more readily comparable of the other two to our models. For example,
our spherical model from Paper~1 would have a variability index of $\sim$1.5 on the
Brand et al. scale.

\subsection{Variation of Viewing Plane}
\label{ss:orbit}

The images discussed in Section~\ref{ss:formal} correspond to a set of positions for a chosen
observer's plane. This plane should not be regarded as typical, so maser ligthcurves were
plotted for a number of different observer's planes (or rotation axes) to yield some statistical information,
and suggest what an `average' observer might record, rather than a very privileged one, who
happens to view the domain rotating close to one of its principal axes.  To this end, all three of the domains plotted
in Fig.~\ref{f:maps} were viewed from 1000 randomly oriented orbital planes, each
perpendicular to an associated rotation axis. Note that in the models
studied in the present work, the observer's plane is always defined by the rotation axis, and
oriented perpendicular to this axis: this is a consequence of converting from the rotating frame
of the cloud, in which the population solution is calculated, to the fixed frame of the observer, in
which the cloud is rotating, with appropriate Doppler corrections to the received frequencies.

\section{Lightcurves}
\label{s:lightcurve}

The principal results of the formal solutions, all of which simulate rotation in the
present work, are lightcurves of maser peak specific intensity as a function
of time. We define the peak specific intensity as the largest specific intensity found
in a particular intensity file, where the intensity is a function of sky position and frequency.
This parameter is comparable to the intensity of the brightest pixel found in any channel
of an interferometer image cube from a
VLBI observation.
The time-base has twice the resolution of that in Paper~1: a 10-year rotation period is divided
into 200 samples, each corresponding to a particular angle around one of the chosen
orbital planes. As in Paper~1, all rotation is solid-body, and stability criteria
strongly suggest that physical changes to the cloud imply that there is no point in
following the rotation for more than one period; these models are most unlikely to
be appropriate for periodic maser flares.

\subsection{Prolate Cloud}
\label{ss:prolate}

\begin{figure}
  \includegraphics[bb=70 50 200 300, scale=0.74,angle=0]{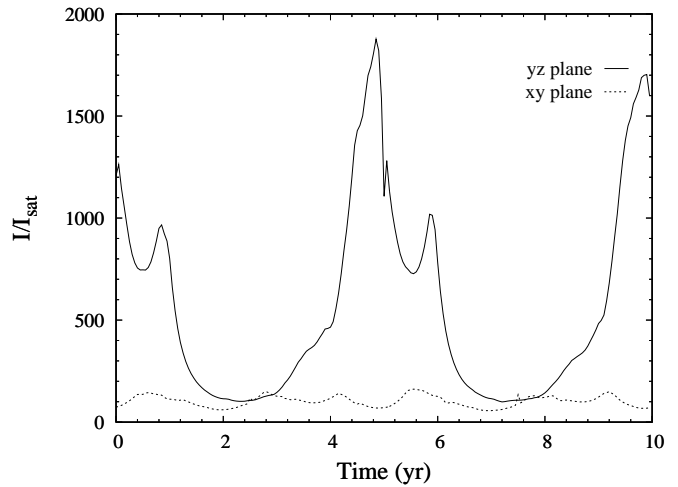}
  \caption{Light curves of the maser emission produced by the rotating prolate cloud 
of deformation factor 0.3 in two
extreme cases are shown: the solid curve results when the cloud is viewed from the $yz$-plane
(rotation about the $x$-axis), and the dashed curve results when the observer's plane
is the $xy$-plane (rotation about the $z$-axis. The optical depth multiplier is 13.0.}
\label{f:prolight}
\end{figure}

\begin{figure}
  \includegraphics[bb=70 50 200 300, scale=0.74,angle=0]{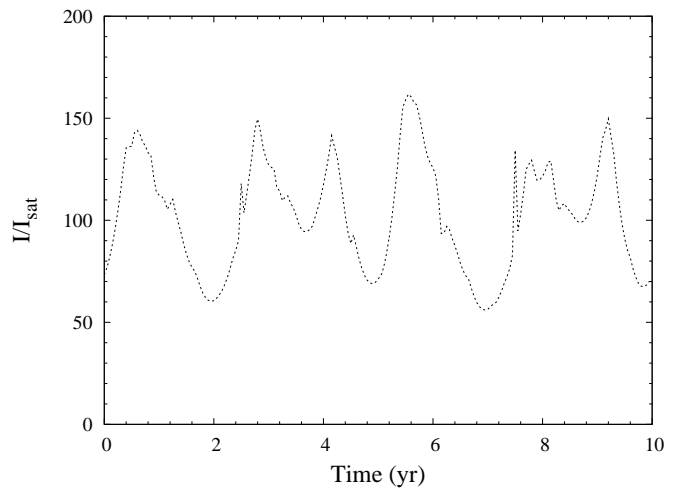}
  \caption{A magnification of the dashed ($xy$-plane) curve from Fig.~\ref{f:prolight}, showing
the detailed variability.}
\label{f:proshort}
\end{figure}

We begin by considering the prolate cloud, and within this we choose conditions likely to
produce extremes of variation. First we consider rotation about the $x$-axis, giving the
$yz$-plane as the observer's plane, as in Fig.~\ref{f:maps}. The long axis of the domain 
lies close to the $yz$-plane, and so, during one rotation, we expect to see two observer's
positions that would produce a specific intensity close to the largest possible value
(see definition in Section~\ref{s:lightcurve} above).
For the $yz$-plane, the
light curve generated by cloud rotation is shown as the solid curve in Fig.~\ref{f:prolight},
where the peak specific intensity rises to a maximum of
1877 times the saturation intensity at approximately half-way through the rotation. The lowest
brightness found was 99.0 at position 144 of 200 (7.2\,yr on the time base). The variability
index is therefore 1877/99 $\simeq$ 19, compared to approximately 3 for the pseudo-spherical
cloud in Paper~1.
By contrast, the second line in Fig.~\ref{f:prolight} (dashed) is the light-curve 
produced when the observer's plane is the $xy$-plane, and
rotation is about the $z$-axis. In this case, the observer is
always viewing the cloud in a projection that is geometrically and optically close
to the thinnest possible. Unsurprisingly, the peak specific intensities are 
generally lower than those following the solid curve (brightest
value of 162), and the variability
index is also substantially lower ($162/56 \simeq 2.89$), since the cloud always presents an
approximately constant range of thicknesses to the observer in this plane. This
value of the variability index is consistent with that found for the pseudo-spherical
object in Paper~1. Two
possible observer's views from the $xy$-plane appear in the top and bottom
left-hand panels of Fig.~\ref{f:domains}.
On the $y$-axis scale of Figure~\ref{f:prolight}, the
variability of the dashed curve is so weak that it is not possible to see
much detail, so this light curve is re-plotted in Fig.~\ref{f:proshort} over a
suitably smaller intensity range.
The light curve in Fig.~\ref{f:proshort} contains many peaks an troughs, 
that reflect the small-scale irregularity of the domain, rather
than a strong pair that correspond to repeated presentation of the long axis towards
the observer. There are no
clear high or low states and no strong regularity in time. This indicates,
as expected, that there are no strongly favoured viewpoints on this observer's plane. 

\begin{figure}
  \includegraphics[bb=70 50 200 300, scale=0.74,angle=0]{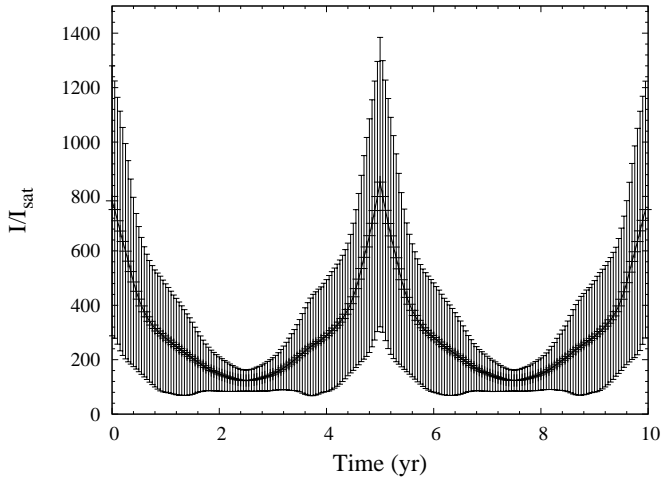}
  \caption{The average light curve resulting from the mean of 1000 rotations about
randomly oriented axes for the prolate domain
with deformation $\Gamma = 0.3$ and optical depth
multiplier of 13.0. Error bars indicate the sample standard deviation at each
time.}
\label{f:prorandom}
\end{figure}

We now abandon the priveleged viewpoints that correspond to the light curves displayed in
Fig.~\ref{f:prolight} and Fig.~\ref{f:proshort}, and consider instead an average light
curve that might correspond to a typical observer. To this end, formal solutions were
computed for 1000 different, randomly oriented, rotation axes, each with 200 observer's
positions that correspond to time snapshots along the light curve. Note that in the
present work, the rotation axes are truly random in space, whilst the similar work
in Paper~1 used observer's planes that all contained the long axis. The resulting
light curves in the present work were then wrapped in time, so that in 
every case the highest intensity found
was placed exactly half way along the time base, at 5\,yr. The final 
light curve, shown in Fig.~\ref{f:prorandom} is the mean over the 1000 random orientations after
wrapping the time base; error bars on this curve indicate
the sample standard deviation in the peak intensity. Unlike Paper~1, we plot only this
one observable.
The variability index of the averaged light curve is 6.92$\pm$0.15: considerably lower than
the extreme variability case in Fig.~\ref{f:prolight}, but larger than the spherical
cloud in Paper~1, or the control example in the present work that has a variability
index of 2.06$\pm$0.02.

\subsection{Oblate Cloud}
\label{ss:oblate}

\begin{figure}
  \includegraphics[bb=70 50 200 300, scale=0.74,angle=0]{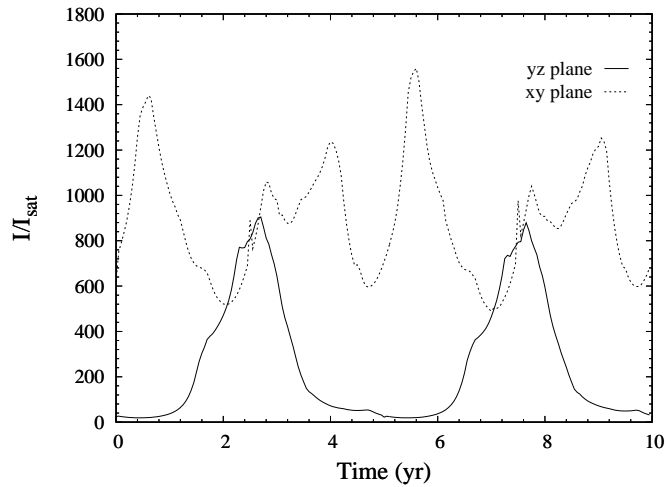}
  \caption{Light curves for an oblate maser cloud 
of deformation factor $-0.30$ rotating about the $x$-axis (solid curve)
and about the $z$-axis (dashed curve). In both cases, the optical depth multiplier is 13.0.}
\label{f:oblight}
\end{figure}

The oblate example cloud has a deformation factor of $-0.30$, so the magnitude is the same as
for the prolate object consisdered in Section~\ref{ss:prolate}. The level of saturation is also
the same (optical depth multiplier equal to 13.00). For an oblate cloud, any observer's plane
in which the object is presented edge-on to the observer is close to
the extreme variability case,
in the sense of the variability index defined in eq.(\ref{eq:appA1}). We chose to use
the same rotation axis (the $x$-axis) as used in the prolate case. The observer's plane
is then the $yz$-plane, and example views include those shown in the right-hand column
of Fig.~\ref{f:maps}.
Rotation about this axis produces repeated edge-on and face-on views of the domain to the
observer, and is a candidate for flaring events. The light curve is shown as the solid
curve in Fig.~\ref{f:oblight}. The variability index is equal to 48.0.
We compare this to a case where the rotation axis is
the $z$-axis (dashed curve in Fig.~\ref{f:oblight}), and the observer's
view, from the $xy$-plane,
is always close to edge-on. In the latter case, the maser emission has high intensity,
but is only weakly variable, according to the variability index defined in eq.(\ref{eq:appA1}).
The actual variability index in this case was 3.19. This value perhaps demonstrates the
weakness of our simple definition of the variability index: although the index for the
dashed curve is much lower on the basis of a maximum to minimum ratio, this curve is
just as variable in terms of maximum to minimum difference.
Moreover, in the case of rotation about the shorter ($z$) axis, an axis of length
similar to the long axis of the domain is always pointing at the observer, whilst
rotation about the $x$-axis only presents an axis of approximately this length towards the observer
twice per rotation. This means that we expect the peaks of the solid curve to be only of similar
intensity to the average intensity of the dashed light-curve (resulting from 
rotation about the $z$-axis). 
This is exactly what we see in Fig.~\ref{f:oblight}. It should be possible to get
a slightly brighter, highly variable, light curve by rotating about an axis exactly
perpendicular to the long axis, but still close to the $xy$-plane, so that
the long axis then does get presented half-periodically to the observer. However,
even this latter case would not be expected to produce peaks of a significantly higher intensity
than those of the dashed curve.

\begin{figure}
  \includegraphics[bb=70 50 200 300, scale=0.74,angle=0]{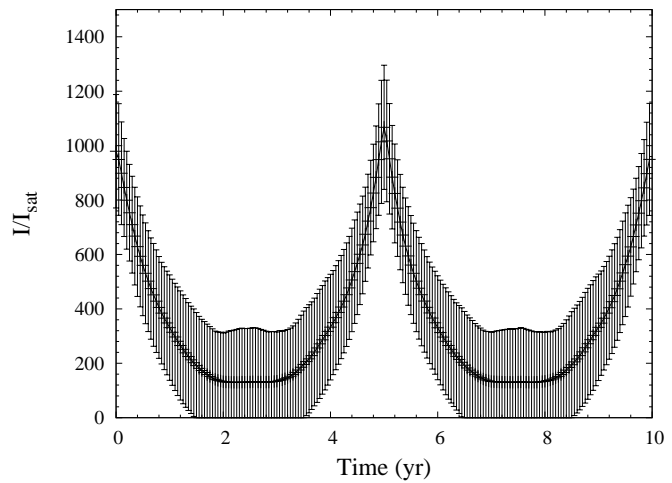}
  \caption{As for Fig.~\ref{f:prorandom}, but for the oblate domain with deformation
factor $\Gamma = -0.3$. The $y$-axis scale is the same as in Fig.~\ref{f:prorandom} for
easy comparison.}
\label{f:oblrandom}
\end{figure}

As in the prolate case, we proceed by considering the case of the light curve that
would be observed by a more typical observer. As in Section~\ref{ss:prolate} this
is based on the average of 1000 observer's planes, and consequently on the same number of
randomly generated rotation axes. The resulting averaged light curve is shown in
Fig.~\ref{f:oblrandom}. Individual light curves were wrapped in time to place the
maximum at 5\,yr before averaging, also as in Section~\ref{ss:prolate}.
Note that the height of the maximum is slightly higher than in the prolate case
(see Fig.~\ref{f:prorandom}) and that the largest standard deviations are now
concentrated towards the fainter parts of the light curve. The variability index
is $8.22\pm0.39$ so, at least on the basis of this very simple statistic, the oblate
domain yields greater variability on average.

\subsection{Effect of Saturation}
\label{ss:satef}

To avoid viewing thousands of light-curves, further work will be based on some
statistics that summarize most of the important information contained
in the data. We use our definition
of the variability index, as defined in eq.(\ref{eq:appA1}), and we also
calculate the duty cycle, defined as the fraction of the full time base
in which the intensity is above half the maximum found.
Unless stated otherwise,
these statistics were calculated from data based on 1000 observer's
planes, as in Fig.~\ref{f:prorandom} and Fig.~\ref{f:oblrandom}. 
Both quantities are calculated with a standard error that is plotted as a $y$-axis
error bar in the relevant figures below.

\begin{figure}
  \includegraphics[bb=70 50 200 300, scale=0.74,angle=0]{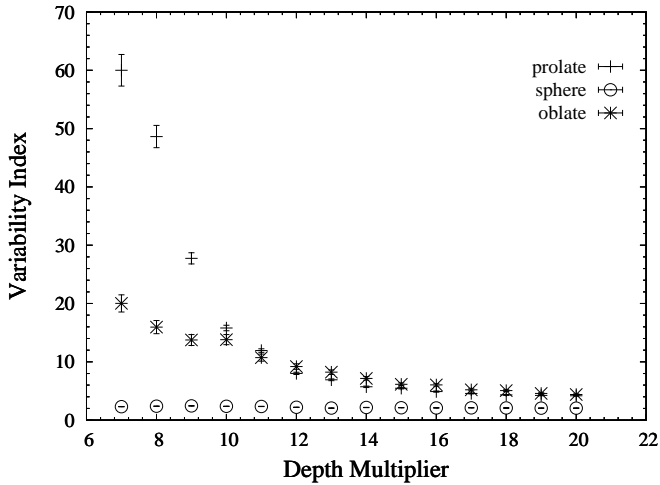}
  \caption{The variability index (ratio of maximum to minimum peak brightness) as a function of
optical depth multiplier for prolate, pseudo-spherical and oblate domains. Error bars show
the standard error based on the 1000 realisations.}
\label{f:average_vindex}
\end{figure}

In this section, we consider the standard prolate and oblate domains with $|\Gamma | = 0.3$
and the pseudo-spherical control domain, but we now vary the degree of saturation by
varying the optical depth multiplier from 7.0 (almost unsaturated, but
still with an amplification factor of order 10$^6$) to 20.0 (strongly saturated).
Results for the average variability index (mean of 1000 light curves, based on randomly
chosen rotation axes) is shown in Fig.~\ref{f:average_vindex}.
The key result is that, for the variability index as defined in eq.(\ref{eq:appA1}), increasing
saturation reduces the index, eventually to a level similar to, but still larger than, the
value of 2-3 typical of the pseudo-spherical domain. However, it should be noted that the
actual maximum brightnesses at $\tau=20$ are of order 3000\,I$_{sat}$, approximately 100 times
larger than at $\tau=7$. From the observer's point of view, it might therefore be more
useful to consider a variability index based on the difference between maximum and minimum
light, rather than the ratio.

\begin{figure}
  \includegraphics[bb=70 50 200 300, scale=0.74,angle=0]{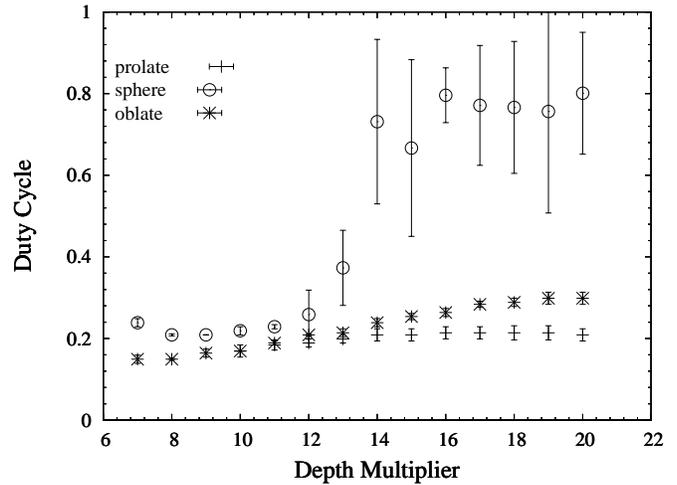}
  \caption{The duty cycle (fraction of time for which the brightness is above
half the maximum value) as a function of
optical depth multiplier for prolate, pseudo-spherical and oblate domains.}
\label{f:average_duty}
\end{figure}

In Fig.~\ref{f:average_duty} we show the second statistic, the duty cycle. 
As in Fig.~\ref{f:average_vindex},
this is based on the average of 1000 realisations. The curves plotted are for the pseudo-spherical
domain and the prolate and oblate clouds with $|\Gamma|=0.3$.
Increasing saturation increases the duty cycle for all cloud types, and the effect
is somewhat stronger for oblate than prolate domains, but much smaller than for the
quasi-spherical case that goes through a rapid switch from low to high duty cycle
at a depth multiplier of $\tau \sim 13$.


\subsection{Effect of the Deformation Factor}
\label{ss:defo}

\begin{figure}
  \includegraphics[bb=70 50 200 300, scale=0.74,angle=0]{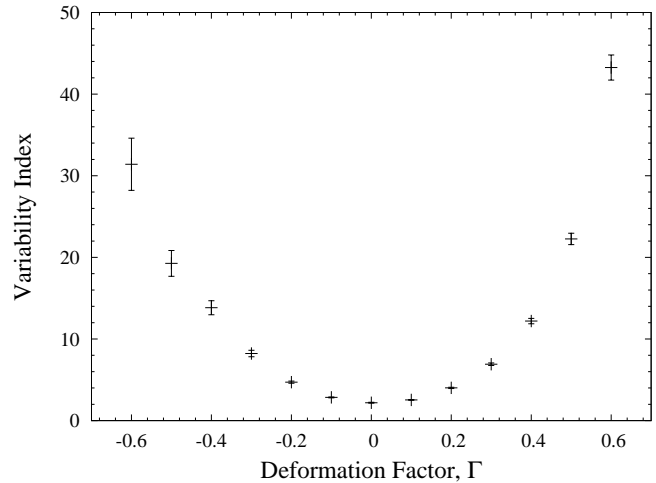}
  \caption{The variability index (ratio of maximum to minumum brightness) as a function of
cloud deformation factor, $\Gamma$. Clouds become incresingly prolate with
increasing $\Gamma$. Error bars show
the standard error based on the 1000 realisations, and all models were run with an optical
depth multiplier of 13.0.}
\label{f:defo_vindex}
\end{figure}

\begin{figure}
  \includegraphics[bb=70 50 200 300, scale=0.74,angle=0]{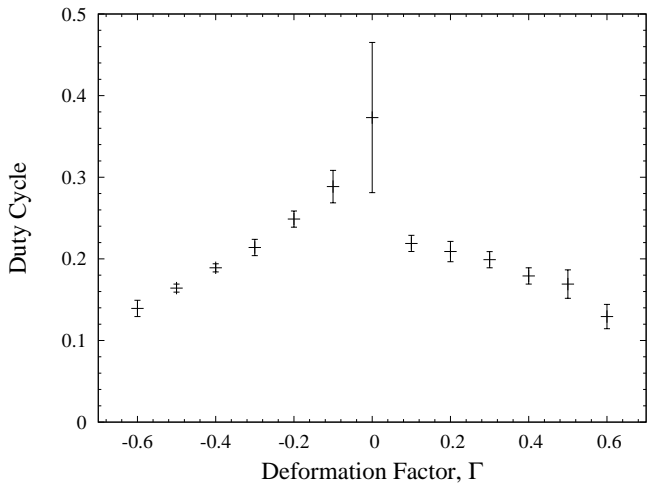}
  \caption{The duty cycle (see caption to Fig.~\ref{f:defo_vindex}) as a function of
cloud deformation factor, $\Gamma$.}
\label{f:defo_duty}
\end{figure}

In this section, we perform calculations similar to those in Section~\ref{ss:satef}, but
now we hold the optical depth multiplier at the standard value of $\tau=13.0$ whilst varying
the deformation factor of the domain. We consider domains from $\Gamma = -0.6$ (extreme
oblate) to $\Gamma = 0.6$ (extreme prolate). The spherical example has $\Gamma = 0$. Results
are shown in Fig.~\ref{f:defo_vindex} for the variability index, averaged over 1000 randomly
selected rotation axes.
We also show the effect of the deformation factor on the duty cycle in Fig.~\ref{f:defo_duty}.
The duty cycle is a strong function of the deformation factor when the magnitude of
this parameter is small. However, the behaviour is quite symmetric for the prolate
and oblate domains, so the duty cycle is not a good statistic for distinguishing
prolate and oblate clouds of the same $|\Gamma|$ observationally.

\subsection{Execution Times}
\label{timings}

Calculation of the nodal solutions still dominates the execution time,
as in Paper~1. However, adoption of the Orthomin($K$) algorithm (see
Section~\ref{s:model}) has reduced this time significantly, particularly at
optical depth multipliers greater than $\sim$12. The new algorithm suffers
from only a weak degredation in performance when strong saturation effects
appear, and the code has been run to optical depth multipliers $>$30.

Under the DiRAC seed-corn grant dp110, better parallelisation of the code, and
other efficiency improvements, have reduced the wall execution time of the version
used in the present work to 28\,s for a single iteration job on a desktop machine with twelve 
Intel Core i7-3930K CPUs, each rated at 3.20\,GHz. This job requires initial
reading of the domain data and ray route tracing in addition to the Orthomin
solution. The same job run on the DiRAC DIAL machine at the University of 
Leicester took 12\,s walltime on the development queue. The code on the DiRAC machine was
compiled with the proprietory Intel compiler, whilst the desktop machine used
the open-source GNU compiler.

\section{Discussion}
\label{discuss}

It is clear that both prolate and oblate maser clouds can produce flaring, and that
the variability index, as judged from any of the definitions in Appendix~\ref{a:indices}, can
be significantly larger than for a pseudo-spherical cloud, given a favourable orientation
of the rotation axis to the observer's line of sight. As expected, the greatest variability
occurs when the long axis of the domain, and an axis similar to the shortest possible,
are sequentially presented to the observer. For a given depth multiplier, the greatest
value of the variability index relates to a prolate domain with its long axis pointing
directly at the observer. However, such an orientation is comparatively rare and, for
a population of clouds with the same magnitude of deformation factor
viewed at random angles, the variability index is comparable for oblate and prolate
clouds. In fact, Fig.~\ref{f:defo_vindex} shows a cross-over between $|\Gamma|=0.4$
and $0.5$, with oblate clouds having a higher variability index at lower values
of $|\Gamma|$ and prolate clouds, at higher values.

From Fig.~\ref{f:average_vindex}, we see that variability index is a decreasing
function of optical depth multiplier. However, what is 
obvious, but not shown in this figure, is
that the maximum brightness is a rising function of depth multiplier, particularly
over the range of $7-12$, where saturation becomes increasingly important. Therefore,
there is likely to be a substantial population of weakly saturated objects with
high variability index that are only marginally detectable. For example an
oblate object with $\Gamma=-0.3$ and depth multiplier of 7 (the lowest value
plotted) has a maximum brightness of only 0.0053 times that of a similar object
at depth 20.0. It is difficult to draw further conclusions without knowledge of
how clouds are distributed in size and/or density in a particular source. However,
for saturated clouds, high variability index (say $\gtrsim$10) requires clouds of
large eccentricity (deformation factor magnitude).

The present work uses the peak specific intensity (see Section~\ref{s:lightcurve}) as
the main measure of maser output, and is therefore more closely related to interferometric
observations than to single-dish spectra. However, spectroscopic counterparts to the
data plotted in Fig.~\ref{f:prorandom} and Fig.~\ref{f:oblrandom}, also averaged over the 1000 realisations,
are available, and
we briefly consider the behaviour of a width parameter, defined as the ratio of the integrated flux
to the peak flux density found in the spectrum. Measured in Doppler units, this
width varied from 0.74 to 0.88 in the oblate cloud of $\Gamma=-0.3$ with a depth multiplier
of 7.0. The prolate cloud of the same $|\Gamma|$ at this depth varied over the larger range
of 0.68 to 0.93. However, for both cloud types, the width is anti-correlated with the flux
density, as would be expected under the CVR approximation, where increased saturation does
not lead to spectral re-broadening. At the higher depth multiplier of 13.0, both the typical
width and its range of variation were reduced (oblate 0.593 to 0.637; prolate 0.596 to 0.640)
but minimum width still correlated with maximum light.

The analysis in Section~\ref{s:lightcurve} offers some hope of distinguishing the
variability of oblate and prolate clouds observationally. If a distribution of similar
objects is present in a source, then a comparison of Fig.~\ref{f:prorandom} and
Fig.~\ref{f:oblrandom} shows that, near the peak of the averaged light curve, the
standard deviation of the maximum brightness is significantly greater in the
prolate case. It is probably best to concentrate on this part of the curve, as it
is the most likely to be observed. However, the bias in the standard deviation is
reversed in the troughs of the averaged light curves. There is also a significant kink
in the wings of the peak in the prolate case (see Fig.~\ref{f:prorandom}). 
Duty cycle is generally higher for
oblate clouds, particularly at large optical depths (see Fig.~\ref{f:average_duty}). For example, any flaring due
to rotation of a $|\Gamma|=0.3$ cloud with a duty cycle above 0.2 is almost certainly
oblate and strongly saturating. 
It is more difficult to say anything about the light curve of a single object,
but a high brightness object with a low variability index is likely to be oblate
and rotating close to edge-on with respect to the observer: in this orientation,
all views of the object over one period have a similar geometrical depth.

Since cloud rotation is just one of many possible mechanisms for generating maser
flares, the aim is to consider several more, and to present similar observational
statistics by which mechanisms may be distinguished. One feature that cannot be
explained by rotation, at least for a cloud that does not distort too much over
a rotational period, is strong asymmetry in the lightcurve, particularly of a type
that yields a longer time for the decay than for the rise on average. Preliminary studies indicate
that superimposition of multiple clouds in the line of sight can produce significantly
smaller duty cycles than the smallest found in Fig.~\ref{f:average_duty} or Fig.~\ref{f:defo_duty},
as well as significantly larger values of the variability index.
However, we leave the details of classification methods to a future publication.

In a little more detail, Fig.~\ref{f:average_vindex} and Fig.~\ref{f:defo_vindex} of the present work suggest that
flares from cloud rotation are unlikely to produce a variability index in excess
of 100. Timescales can be estimated from the simple stability criterion discussed
in Paper~1. If the maximum rotation rate consistent with stability over one period
is taken to be $\Omega_{max} = (2\beta /R)^{1/2}$ for pressure parameter $\beta$ and cloud
radius, $R$, then the corresponding minimum period, or timescale more generally, is
\begin{equation}
P_{min} = 2.18 \times 10^3 \left[
                            \frac{\mu_H R_{AU}}
                                 {\alpha_{-3} n_6 \sigma_{-20} T_3}
                         \right]^{1/2} \;\;{\rm d},
\label{eq:pmin}
\end{equation}
where $\mu_H$ is the mean molecular mass of the gas external to the cloud in 
hydrogen atom units, $R_{AU}$ is the cloud radius in astronomical units,
$\alpha_{-3}$ is in units of 10$^{-3}$, $\sigma_{-20}$ is the collision cross-section
in units of 10$^{-20}$\,m$^2$, $n_6$ is the number density of the external gas in
units of cm$^{-3}$, and $T_3$, the temperature of the external gas in units of 1000\,K.
Allowing for favourable values of the parameters in eq.(\ref{eq:pmin}), we should
therefore expect timescales of order a few years or more in addition to modest
values of the variability index. Rotation can therefore already be ruled out
as a mechanism for flares with timescales of weeks to months.

\section{Conclusions}
\label{conclusion}

Rotation of irregular, approximately spheroidal, clouds provides a mechanism for maser
flares of modest variability index and long timescale. The variability index ranges from
a few to $\lesssim$100 if the index is the one
defined in eq.(\ref{eq:appA1}). Shorter timescales than $\sim$1\,yr are excluded on
the grounds of stability of the cloud, as are truly periodic flares. Given these
restrictions, rotation is more suited to possibly providing some of the variability
associated with Class~II methanol masers than with very bright and rapid H$_2$O maser
flares.

Both prolate and oblate objects can produce strong flares. The averaged light curves of the
two types are quite strong functions of the degree of saturation in the cloud and of
the degree of deformation, with more extreme objects exhibiting stronger variability.
Variability declines with increasing saturation for an object of given eccentricity, but
the highly variable, weakly saturating objects are typically hundreds of times less
bright, and so less likely to be visible. Detailed differences in the average behaviour
of well-chosen statistics may enable us to distinguish between the maser emission
from oblate and prolate clouds.

The choice of variability index used in this work is arguably not the best possible,
and some discussion over the most useful definition of this index
for maser variability studies is desirable.

\section*{Acknowledgments}

MDG and SE acknowledge funding from the UK Science and Technology Facilities
Council (STFC) as part of the consolidated grant ST/P000649/1 to the Jodrell Bank
Centre for Astrophysics at the University of Manchester. 
This work was performed, in part, using the DiRAC Data Intensive service at Leicester, operated 
by the University of Leicester IT Services, which forms part of the STFC 
DiRAC HPC Facility (www.dirac.ac.uk). The equipment was funded by BEIS capital 
funding via STFC capital grants ST/K000373/1 and ST/R002363/1 and 
STFC DiRAC Operations grant ST/R001014/1. DiRAC is part of the National e-Infrastructure.
In particular, the authors would like
to thank the DiRAC software engineers, Jon Wakelin and Samuel Cox for their
contributions to the much improved code performance in this work under seedcorn
grant dp110. 

\bibliographystyle{mn2e}
\bibliography{MDGray_MN19}

\appendix

\section[]{Variability Indices}
\label{a:indices}

Unfortunately, whilst any study of maser variability needs some measure of the strength
of that variability, there does not at present seem to be any accepted standard for
such a measure, or index. In Section~\ref{ss:formal}, three different variability indices
were introduced, all of which are defined below. In Paper~1 we used the index,
\begin{equation}
{\cal V}_1 = S_{max}/S_{min} ,
\label{eq:appA1}
\end{equation}
where $S$ is a flux or intensity measure, and $S_{max}$ and $S_{min}$ denote the
respective  maximum and minimum values of that measure found in a given time
interval. In Paper~1, three different quantites were used for $S$: the maximum
pixel brightness (specific intensity) found in an image; the flux density in the central
spectral channel and the frequency-integrated flux. For simplicity, only the first
of these measures is used for $S$ in the present work.

In their study of 22-GHz H$_2$O maser variability, \citet{2007IAUS..242..223B} use
the variability index,
\begin{equation}
{\cal V}_B = F_{max} / \bar{F} ,
\label{eq:appA2}
\end{equation}
where $F$ is the flux density (in a specified channel) and $\bar{F}$ is its mean value
over the duration of the observing programme.

In their extensive variability survey of 6.7\,GHz methanol masers,
\citet{2004MNRAS.355..553G} use the variability index definition, calculated for
a specific velocity or frequency channel, of
\begin{equation}
{\cal V}_G = \frac{\sum_{j=1}^J (F(t_j)-\bar{F})^2 - \sum_{j=1}^J (F_N(t_j)-\bar{F}_N)^2}
                  {\bar{F}^2},
\label{eq:appA3}
\end{equation}
where $t_j$ is the time of the $j$th sample in a series of $J$ observations. $F(t_j)$
is the flux density at time $t_j$ and $\bar{F}$ is the mean flux density over
all $J$ observations. Quantities with the subscript $N$ are calculated from a
noise channel, which implies a channel without maser emission. This definition
was adopted from earlier work on Cepheid variables by \citet{1996PASP..108..851S}.

The present authors have taken the liberty of squaring the mean flux density in the denominator
of eq.(\ref{eq:appA3}) in order to yield a dimensionless variability index: the
original in equation~1 of \citet{2004MNRAS.355..553G} has units of flux density as
written, but the index does appear dimensionless in their tables, and it is also
in \citet{1996PASP..108..851S}, though with a different normalization.

\end{document}